\documentclass[aps,pra,reprint,showpacs,nofootinbib,superscriptaddress]{revtex4-1}
\input{Qcircuit}
\usepackage{graphicx,comment}
\usepackage{amsfonts}
\usepackage{amsmath,amssymb}
\usepackage{bm}
\usepackage{color}
\usepackage{epstopdf}
\usepackage{epsfig}
\usepackage[font=small,
   justification=justified,
   format=plain]{caption} 
\usepackage{subfig}
\usepackage{verbatim}
\usepackage{hyperref}
\usepackage{multirow}
\usepackage[table,xcdraw]{xcolor}
\usepackage{tabularx}
\usepackage{wrapfig}
\hypersetup{
     colorlinks   = true,
     citecolor    = blue
}
\usepackage{lineno}
\usepackage[thicklines]{cancel}
\usepackage{color, colortbl}
\definecolor{LightCyan}{rgb}{0.88,1,1}
\usepackage{url}
    
{\rm }

\def\be{\begin{equation}}
 \def\ee{\end{equation}}
 \def\bea{\begin{eqnarray}}
 \def\eea{\end{eqnarray}}

\def\2{\frac{1}{2}}
\def\4{\frac{1}{4}}

\newcommand{\expect}[1]{\langle {#1} \rangle}

\begin{document}

\footnotetext{This manuscript has been authored by UT-Battelle, LLC, under Contract No. DE-AC0500OR22725 with the U.S. Department of Energy. The United States Government retains and the publisher, by accepting the article for publication, acknowledges that the United States Government retains a non-exclusive, paid-up, irrevocable, world-wide license to publish or reproduce the published form of this manuscript, or allow others to do so, for the United States Government purposes. The Department of Energy will provide public access to these results of federally sponsored research in accordance with the DOE Public Access Plan.}%

\title{Practical Quantum Computation of Chemical and Nuclear Energy Levels Using Quantum Imaginary Time Evolution and Lanczos Algorithms}

\author{K\"ubra Yeter-Aydeniz }
\email{yeteraydenik@ornl.gov}
\affiliation{Physics Division, Oak Ridge National Laboratory,
  Oak Ridge, TN 37831, USA}

\author{Raphael C.\ Pooser}
\email{pooserrc@ornl.gov}
\affiliation{Computational Sciences and Engineering Division, Oak Ridge National Laboratory,
  Oak Ridge, TN 37831, USA}

\author{George Siopsis}
\email{siopsis@tennessee.edu}
\affiliation{Department of Physics and Astronomy,  The University of Tennessee, Knoxville, TN 37996-1200, USA}

\date{\today}

\begin{abstract}
Various methods have been developed for the quantum computation of the ground and excited states of physical and chemical systems, but many of them require either large numbers of ancilla qubits or high-dimensional optimization. The quantum imaginary-time evolution (QITE) and quantum Lanczos (QLanczos) methods proposed in \cite{Motta2019} eschew the aforementioned issues. In this study, we demonstrate the practical application of these algorithms to nontrivial quantum computation, using the deuteron binding energy and molecular Hydrogen binding and excited state energies as examples. With the correct choice of initial and final states, we show that the number of time steps in QITE and QLanczos can be reduced significantly, which commensurately simplifies the required quantum circuit and improves compatibility with NISQ devices. We have performed these calculations on cloud-accessible IBM-Q quantum computers. With the application of readout-error mitigation and Richardson error extrapolation, we have obtained ground and excited state energies that agree well with exact results obtained from diagonalization.

\end{abstract}

\maketitle

\section{Introduction}
Noisy intermediate scale quantum (NISQ) computers have recently become workhorse platforms for the study of codesign and the design of near-term quantum algorithms.
Thus far, the variational quantum eigensolver has proved to be one of the most useful applications for these devices.
Variational methods have been used to solve problems in chemistry, nuclear physics, quantum field theory, high energy physics, and others~\cite{OMalley2016, linke2017experimental, kandala2017, Dumitrescu2018, klco_quantum-classical_2018, Colless2017, mccaskey_quantum_2019}. 
While these small-scale applications show promise for using NISQ devices to sample from distributions and calculate expectation values, short coherence times make calculations involving time evolution exceedingly difficult on NISQ devices. 
Time evolution calculations hold promise for calculating scattering amplitudes ~\cite{jordan_quantum_2014} and, excited~\cite{mcclean_hybrid_2017,Higgott2019}, and non-equilibrium states~\cite{lamm_simulation_2018}.
One approach to the problem of short coherence times is quantum imaginary time evolution (QITE)~\cite{McArdle2019}, in which non-unitary evolution can be calculated variationally. Combining QITE with the Lanczos optimization method (referred to as QLanczos in the context of quantum computing), one can obtain time-evolved phenomena of various many body systems~\cite{Motta2019}.

Here, we demonstrate the practical application of QITE and QLanczos on current cloud-based NISQ hardware in order to calculate ground and excited states in different fields of study. We use the method to obtain the ground state of the deuteron nucleus in one instance, and we calculate both the ground and excited states of the H$_2$ molecule in another. The quantum computations were done on several cloud-accessible IBM Q Experience devices, i.e. 20-qubit Johannesburg, 20-qubit Poughkeepsie, 53-qubit Rochester, and 5-qubit Yorktown hardware. The results obtained from the quantum computations were compared with the classical calculations obtained from exact diagonalization. Despite the fact that we used a simplified version of the deuteron Hamiltonian we were able to obtain the ground state energy of deuteron without the need for any non-linear optimization or ancillae. We also obtained the energy spectrum of H$_2$ molecule very close and even within chemical accuracy ($1.6 \times 10^{-3}$ Hartree). These demonstrations show great promise for scaling up time evolution as a solution method on near-term quantum hardware, and they illustrate that the approaches have practical, near-term applicability to an array of fields from high energy physics to chemistry.





Quantum imaginary time evolution addresses the problem of exponentially increasing resource requirements for computation as a function of the number of interacting particles.
It replaces the real time in the time-dependent Schr\"odinger equation with imaginary time $(t \to -i \beta)$. The solution to this equation involves an imaginary-time evolution operator, $\mathcal{U}=e^{-\beta H}$. This operator leads to the decay of all states except for the ground state. Therefore, the normalized imaginary-time evolution of a state can be expressed as
\be
|\Psi(\beta)\rangle=\frac{e^{-\beta H}|\Psi(0)\rangle}{||e^{-\beta H}|\Psi(0)\rangle||}~,\label{eq:im_time}
\ee
where $\beta$ is the imaginary time \cite{Magnus1954} and $|| \cdot || \equiv \sqrt{\langle \cdot | \cdot\rangle}$ is the state norm. 

Quantum computation of the ground state energy of many-body systems using the imaginary-time evolution can be thought of as a natural alternative as quantum computers provide exponential speed ups. 
The basic idea behind QITE \cite{McArdle2019} is to approximate the non-unitary imaginary-time evolution in small steps with unitary updates on a set of qubits including data qubits and ancilla qubits. By tracing over the ancilla, the data qubits effectively see non-unitary evolution, which allows us to approximate imaginary time evolution and calculate the decay to the ground state via \eqref{eq:im_time}. However, the algorithm of Motta et al. \cite{Motta2019} eliminates ancillae as a requirement, considerably simplifying the algorithm. On a quantum computer, the unitary evolution utilizes Trotterization. Current quantum computers are incapable of simulating long time evolution, or a large number of Trotter steps, due to short coherence times and excessive gate noise that further reduces coherence time. However, since QITE seeks to approximate non-unitary evolution with a unitary operator, we can reduce the number of Trotter steps by calculating a specific unitary that corresponds to the largest possible steps in imaginary time that yield a given desired accuracy. This amounts to solving a linear system of equations that provide coefficients of expansion, in terms of Pauli operators, for the unitary evolution operators. In the case of the deuteron, we found that solving this system of equations for the largest timesteps provided a unitary evolution operator that corresponded to the familiar unitary coupled cluster (UCC) ansatz \cite{Li2017}.

While this was a large simplification of the QITE algorithm, a key advantage over variational methods is the ability to use the method in a QLanczos algorithm to calculate excited states. The basic idea behind the QLanczos algorithm is to fill the Krylov space with vectors in powers of $e^{-2\Delta \tau H}$, which is done using QITE, and then these vectors are used to calculate Hamiltonian matrix elements, which leads to a generalized eigenvalue equation, yielding a computation of ground and excited states.
Using the \textit{single-step} method in QITE, we reduced the depth of the quantum circuit, which makes these algorithms more compatible with NISQ \cite{Preskill2018quantumcomputingin} devices.

The rest of the paper is organized as follows. 
In section \ref{sec:qcomp} we present how the quantum computations are conducted, including the algorithms (sections \ref{sec:QITE} and \ref{sec:QLanc}). Following this, in section \ref{sec:res} we present the results of our quantum computations conducted using several IBM Q quantum processors.
Finally, in section \ref{sec:con}, we provide a summary and future work. In the appendix we provide information on the model Hamiltonians that we used for deuteron and molecular Hydrogen (section \ref{sec:mod}), error mitigation strategies used (section \ref{sec:err}) and some details of the calculations (section \ref{sec:supp}).

\section{Quantum Computation}
\label{sec:qcomp}
\subsection{Algorithms}
Here, we present a brief review of the QITE and QLanczos algorithms that were proposed in \cite{Motta2019}.


\subsubsection{Quantum Imaginary Time Evolution (QITE)}\label{sec:QITE}

To be able to simulate the dynamics of many-body systems we need to break down the Hamiltonian of these systems into local components such that $H=\sum_m^M h_m$ where $h_m$ are non-commuting local terms of the system \cite{Jones2019}. For many-body systems, the number of terms in the Hamiltonian scales polynomially with the number of particles in the system. For example, the $N=2$ deuteron Hamiltonian in \eqref{DeutHam} can be decomposed into 
\be
\begin{split}
&h_1=5.906709 I + 0.218291 Z_0- 6.125  Z_1 ~,\\& 
h_2=-2.143304(X_0X_1+Y_0Y_1)~.
\end{split}
\ee
Because of the non-commuting terms in the Hamiltonian the decomposition of the evolution into small time steps and decomposing these steps into local gates can be done using the first order Lie-Trotter-Suzuki decomposition formula \cite{Trotter1959} which gives
\be
\mathcal{U}=(\prod_{m=1}^M e^{-\Delta \tau h_m})^n+\mathcal{O}(\Delta \tau)
\ee 
where $n=\frac{\beta}{\Delta \tau}$ is the number of steps in the evolution.


For two non-commuting operators the matrix exponential can be written as
\be
e^{-A\Delta \tau}e^{-B\Delta \tau}=e^{-\left(A+B\right)\Delta \tau-\frac{1}{2}\left[A,B\right] (\Delta \tau)^2+\dots}~,
\ee
following the Baker-Campbell-Hausdorff lemma.

This formula is given for two operators only, but it can be generalized to $n$ operators. In our calculations assuming that $\Delta \tau$ is small we can approximate the imaginary-time evolution up to an order of $\mathcal{O}(\Delta \tau)$ as follows.
\be
|\Psi (\beta)\rangle\approx c_n(e^{-(h_1+h_2+\dots+h_M)\Delta \tau})^n |\Psi(0)\rangle~,
\ee
where \be c_n=\frac{1}{\sqrt{\langle \Psi(0)|(e^{-(h_1+h_2+\dots+h_M)\Delta \tau})^{2n} |\Psi(0)\rangle}} \ee is the normalization constant. 

The $s$-th step of the imaginary-time evolution can be written as
\be
|\Psi_s\rangle = c_s e^{-(h_1+h_2+\dots+h_M)\Delta \tau} |\Psi_{s-1}\rangle~,\label{sthstep}
\ee
where $s=1, 2, \dots, n$.
The purpose of the QITE algorithm is to approximate \eqref{sthstep} with unitary updates such that
\be
|\Psi_s\rangle \approx e^{-i\Delta \tau A_s}|\Psi_{s-1}\rangle~.
\ee
where $A_s$ can be written in terms of Pauli operators (defined in \eqref{eq:Pauli}) up to $D+1$ qubits and can be expressed as
\be
A_s=\sum_{i_0 i_1 \dots i_D}a[s]_{i_0 i_1 \dots i_D}\sigma_{i_0}\sigma_{i_1}\dots \sigma_{i_D}~.\label{As}
\ee
For our two (three)-qubit systems we used $D=1$ ($D=2$).
To be able to approximate the imaginary-time evolution with these unitary updates we need to calculate the coefficients $a[s]$. For small $\Delta \tau$, up to an order of $\mathcal{O}(\Delta \tau)$, the coefficients are found by solving a linear system of equations ${\bm{\mathcal{S}} \bm{a}}[s]={\bm{b}}$ at every step of the imaginary-time evolution,
where
\be
\mathcal{S}_{\mathcal{I},\mathcal{I}'}[s]=\langle \Psi_s|\sigma_{i_0}^\dagger\sigma_{i_1}^\dagger\dots \sigma_{i_D}^\dagger\sigma_{i'_0}\sigma_{i'_1}\dots \sigma_{i'_D}|\Psi_s\rangle~,
\ee
\be
b_{\mathcal{I}}[s]=-ic_s^{-1/2}\langle \Psi_s |\sigma_{i_0}^\dagger\sigma_{i_1}^\dagger\dots \sigma_{i_D}^\dagger h_m|\Psi_s \rangle
\ee
with $\mathcal{I}=i_0,i_1,\dots, i_D$.
The solution to this equation minimizes the operator norm $||c_{s}^{-1/2}|\Psi_{s}\rangle-(1-i\Delta \tau A_s)|\Psi_{s-1}\rangle||$. More detailed discussion on the calculation of the coefficients $a[m]$ can be found in the supplementary information of ref.\ \cite{Motta2019}. 


The calculation of the unitary updates for our deuteron and molecular Hydrogen examples gave us interesting results. For $N=2$ case the unitary updates have the form of $A_s=a[s]\left(X_0 Y_1-X_1Y_0\right)$ and $N=3$ the unitary updates have the form of $A_s=a_1[s](X_0Y_1-X_1Y_0)+a_2[s](X_0Z_1Y_2-X_2Z_1Y_0)$ which are in the same form as UCC (unitary coupled cluster) ans\"atze that were proposed for molecular Hydrogen in \cite{OMalley2016} and deuteron in \cite{Dumitrescu2018}. This means that the unitary updates recover the UCC ansatz. 

Using QITE it is possible to obtain the excited state energies since the system does not necessarily converge to the ground state, but rather depends on the initial state, $|\Psi_0\rangle$, choice. In general, the system converges to the eigenvalue of the Hamiltonian whose eigenvector is non-orthogonal to the initial state, $|\Psi_0\rangle$.

\subsubsection{Quantum Lanczos (QLanczos) Algorithm}\label{sec:QLanc}

The QLanczos algorithm is based on the QITE algorithm, but provides the advantages of faster convergence in special cases, and it can be used to calculate excited state energies. The basic idea behind the QLanczos algorithm is to fill in the Krylov subspace with vectors in powers of $e^{-2\Delta \tau H}$ at each Lanczos iteration such that $\mathcal{K}: \{|\Phi\rangle, e^{-2\Delta H}|\Phi\rangle, e^{-4\Delta H}|\Phi\rangle, \dots\}$. The vectors in the Krylov subspace are obtained using the QITE algorithm as
\be
|\Phi_l\rangle=c_l e^{-l\Delta \tau H}|\Psi_t\rangle
\ee
for $0 \leq l < L_{\text{max}}$ assuming $l$ is an even number. Here, $|\Psi_t\rangle=c_t\left(\prod_{s=1}^t  e^{-i\Delta\tau A_s}\right)|\Psi_0\rangle=|\Phi_0\rangle$ is the initial QLanczos state which is obtained from QITE subroutine.
After building the Krylov subspace we need to calculate the overlap matrix elements $(\mathcal{T}_{l,l'})$ and Hamiltonian matrix elements $(\mathcal{H}_{l,l'})$ in terms of the expectation values since they are the only experimentally accessible values. The calculations give overlap and Hamiltonian matrix elements as
\be
\mathcal{T}_{l,l'}=\langle \Phi_l|\Phi_{l'}\rangle=\frac{c_l c_{l'}}{c_r^2}~,
\ee
\be
\mathcal{H}_{l,l'}=\langle \Phi_l|H|\Phi_{l'}\rangle=\mathcal{T}_{l,l'}\langle \Phi_r|H|\Phi_r\rangle~,
\ee
where $r=\frac{l+l'}{2}$. The normalization constants can be recursively calculated in terms of expectation values using 
\be
\frac{1}{c_{r+1}^2}=\frac{\langle \Phi_r|e^{-2\Delta \tau H}|\Phi_r \rangle}{c_r^2}~.
\ee
The next step of the QLanczos algorithm is to utilize the calculated overlap and Hamiltonian matrix elements and solve the generalized eigenvalue equation
\be
{\bm{\mathcal{H} x}} =E {\bm{\mathcal{T}x}}~. \label{gen_eig}
\ee 
The ground and excited states can then be found from the eigenvectors of the generalized eigenvalue equation. For example, the normalized ground (g) (excited (e)) state approximation is
\be
|\Phi_{\text{g (e)}}\rangle=\frac{\sum_{l=0,2,\dots}^{L_{\text{max}}}{x_l}_{\text{g(e)}}|\Phi_l\rangle}{||\sum_{l=0,2,\dots}^{L_{\text{max}}}{x_l}_\text{g(e)}|\Phi_l\rangle||}~, \label{QLancEq}
\ee
where the coefficients $x_{l_{\text{g(e)}}}$ are obtained from the eigenvector that corresponds to the ground (excited) state energy such that $\begin{pmatrix}{x_0}_{\text{g(e)}}  & {x_2}_{\text{g(e)}} &  \dots & {x_{L_{\text{max}}}}_{\text{g(e)}}\end{pmatrix}^T$. Then the energy expectation values are calculated from 
\be
E_{\text{g (e)}}=\langle \Phi_{\text{g (e)}}|H|\Phi_{\text{g (e)}}\rangle \label{QLancEn}
\ee
which then leads to calculation of the ground and excited state energies using QLanczos algorithm. In the exact calculations the energy values obtained from the eigenvalues of the generalized eigenvalue equation \eqref{gen_eig} match with the values obtained from \eqref{QLancEn}. Our quantum computation shows that using \eqref{QLancEn} is numerically more stable and gives much better results than using the eigenvalues of \eqref{gen_eig} as seen in Table \ref{table:QLanczosDeuteron}.


The QLanczos method converges much faster than the QITE algorithm but one needs to do measurements at each imaginary-time projection of the Krylov subspace vectors to obtain the corresponding overlap and Hamiltonian matrix elements from the expectation values.
The more vectors in the Krylov subspace the more QITE measurements with an increasing quantum circuit depth are required. At this point, the \textit{single-step} method we proposed that is explained in section \ref{sec:QProgram} plays an important role in terms of reducing the circuit depth and possible noise that will arise due to the gates in the circuit. 

\subsection{Quantum Program}
\label{sec:QProgram}
\begin{figure}[ht!]
    \centering
    \includegraphics[scale=0.4]{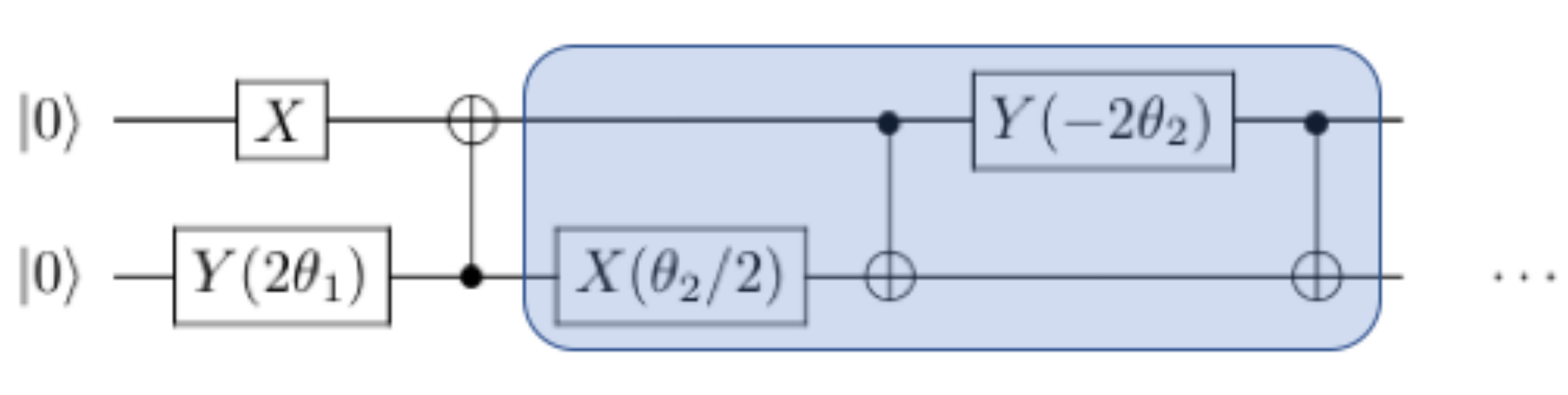}
    \caption{Two-qubit QITE quantum circuit with initial state $|\Psi_0\rangle =|10\rangle$. The quantum circuit in the box is repeated at each QITE step after the second step of the algorithm for convergence. }
    \label{fig:N2DeuteronQCircuit1}
\end{figure}

As mentioned in section \ref{sec:QITE}, the imaginary-time evolution in QITE algorithm is provided by unitary updates of the form 
$
\mathcal{U}_s=e^{-i\Delta \tau \, a[s]\left(X_0 Y_1-X_1 Y_0\right)}
$
for our two-qubit examples. One way to obtain the ground state energy using QITE is to start with an initial product state, say $|\Psi_0\rangle=|10\rangle$ and apply the unitary updates while calculating the coefficients $a[s]$ that give the state $|\Psi_s\rangle $ at every step of the imaginary-time evolution.
At the end of the $n-$th step of the imaginary-time evolution one expects to reach the ground state energy. This version of QITE would require a quantum circuit as seen in Fig. \ref{fig:N2DeuteronQCircuit1}, which only shows the first two steps of the imaginary-time evolution; the depth of the quantum circuit increases as the number of steps increases. 
At every step of the imaginary-time evolution, the quantum circuit in the shaded area is repeated such that $\theta_s=2\Delta \tau a[s]$. Naturally, large depth circuits are very noisy, and not necessarily amenable to error mitigation techniques.  


\begin{figure}[ht!]
    \centering
    \includegraphics[scale=0.4]{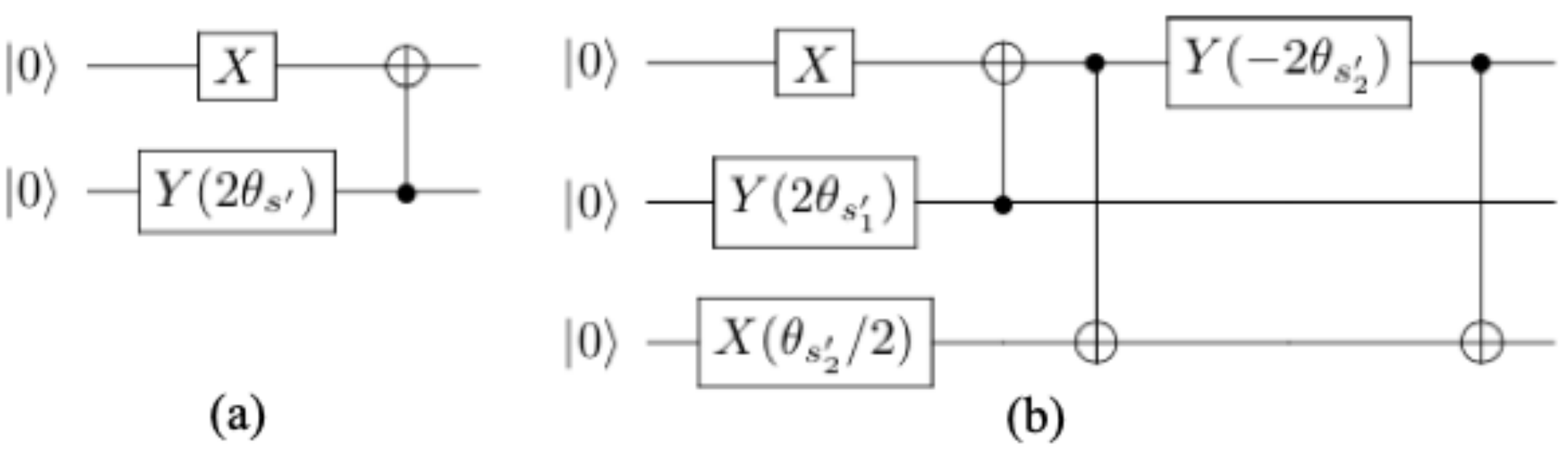}
    \caption{Two (panel (a)) and three (panel (b))-qubit \textit{single-step} QITE quantum circuit with initial state $|\Psi_0\rangle =|10\rangle$ and $(|\Psi_0\rangle=|100\rangle)$, respectively. The angle parameters are calculated for the same circuit until the convergence is reached.}
    \label{fig:DeuteronQCircuit2}
\end{figure}

To make QITE more compatible with NISQ devices we reduce the number of time steps. In the \textit{single step} version, instead of building the quantum circuit that combines each unitary update which gives $|\Psi_{s}\rangle \approx e^{-i\Delta \tau A[s]}|\Psi_{s-1}\rangle$ we build the quantum circuit based on the calculated coefficient $A_s'$ that gives $|\Psi_s\rangle \approx e^{-i\Delta \tau s A_s'} |\Psi_0\rangle$.
In this case, the quantum circuit is given in Fig.~\ref{fig:DeuteronQCircuit2} (a) which only includes one CNOT gate for a specific initial state of $|\Psi_0\rangle=|10\rangle$.
The rotation angle is now defined as $\theta_{s'}=2s\Delta \tau  a'[s]$ such that $\beta'=s\Delta \tau $ is the imaginary-time corresponding to a specific expectation value, and at $\beta=n\Delta \tau$ the energy converges to the ground (or excited) state energy. We run the same quantum circuit with different calculated $a'[s]$ coefficients until the energy expectation value converges to the ground (or excited) state energy. 


Applying the same strategy to our three-qubit deuteron example with an initial state of $|\Psi_0\rangle=|100\rangle$ gives the unitary updates of the form 
\be
\mathcal{U}_{s'}\approx e^{-i\Delta \tau s a_1'[s](X_0Y_1-X_1Y_0)}e^{-i\Delta \tau s a_2'[s](X_0Z_1Y_2-X_2Z_1Y_0)}~.
\ee
with $\theta_{s_i'}=2s\Delta \tau a_i'[s]$ for $i={1,2}$ which can be approximated with the quantum circuit in Fig. \ref{fig:DeuteronQCircuit2}(b).

In addition to our \textit{single-step} QITE approach we also applied the error mitigation strategies to improve results (explained in Sec. \ref{sec:err} in detail). In what follows, we applied these error mitigation strategies to obtain the energy expectation values. 

\subsection{Results and Discussion}
Here, we present the experimental results from IBM Q hardware for QITE and QLanczos algorithms. Information on the experiments and the hardware used can be found in Table \ref{table:Hardware} of the supplementary information.
\subsubsection{QITE results}
\label{sec:res}
\textit{Deuteron}

\begin{figure*}[ht!]
    \centering
    \subfloat[]{\includegraphics[width=0.5\linewidth]{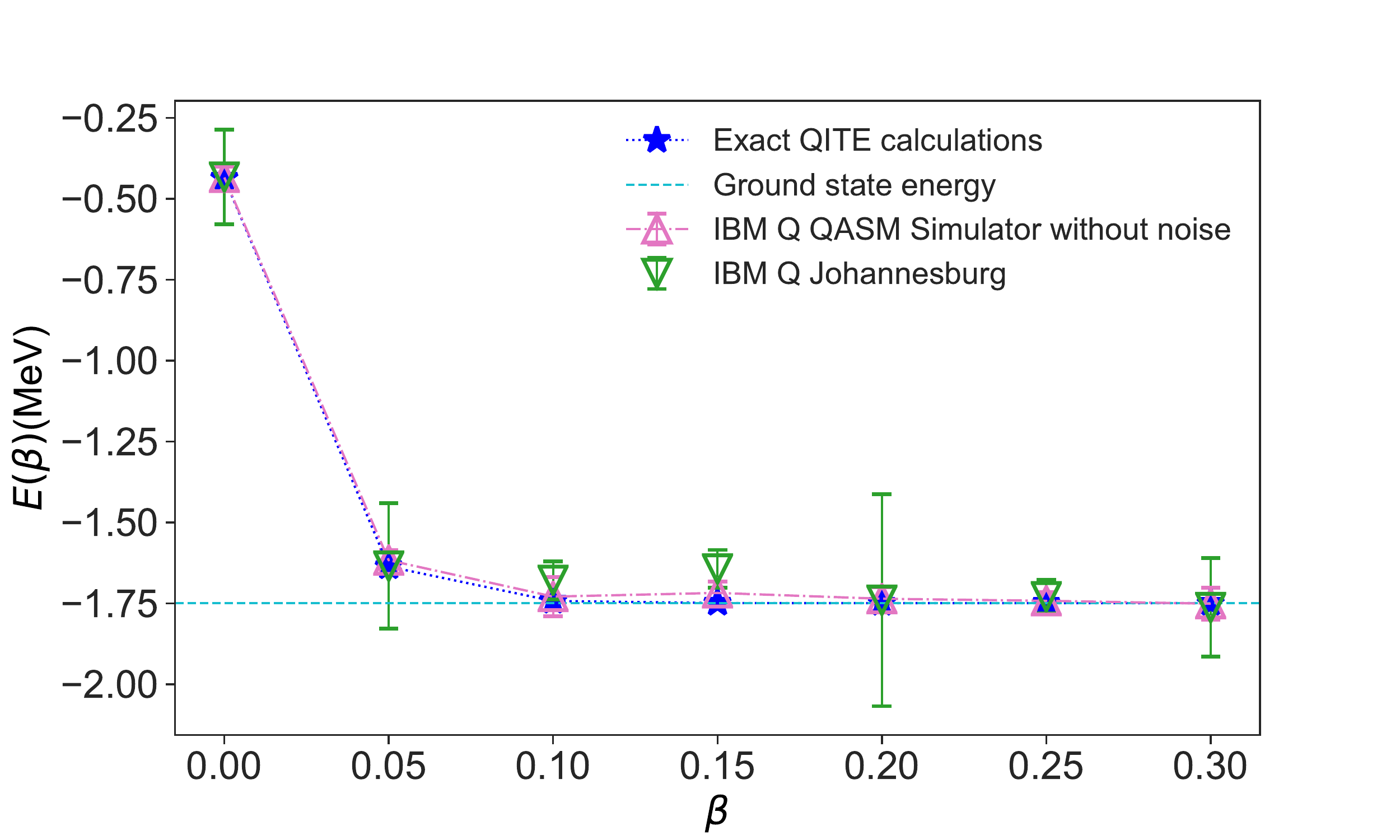}} 
    \subfloat[]{\includegraphics[width=0.5\linewidth]{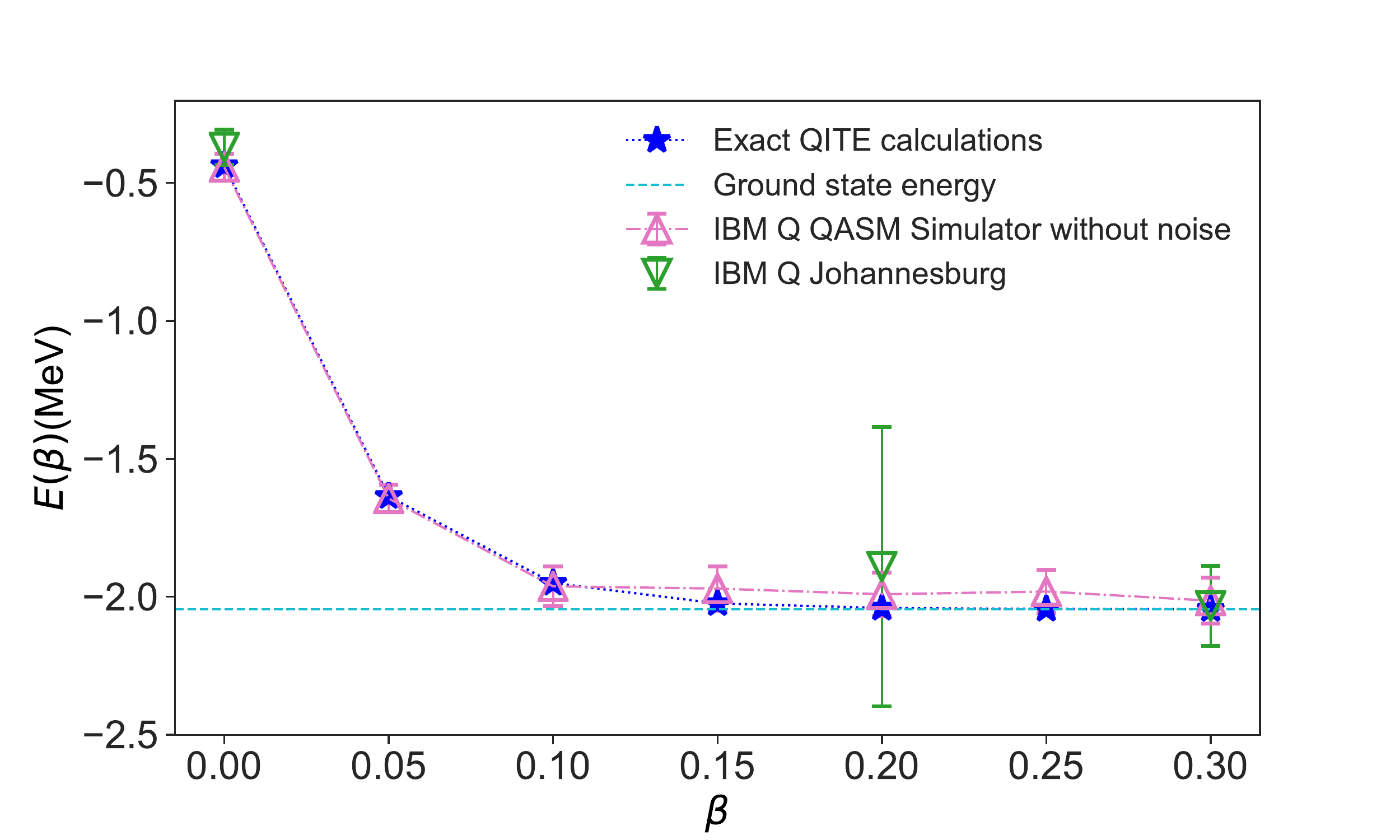}} 
    \caption{(Color online) Energy expectation values of deuteron as a function of imaginary-time for  $N=2$ (panel (a)) with $|\Psi_0\rangle = |10\rangle $ and $N=3$ (panel (b)) with $|\Psi_0\rangle=|100\rangle$. (a) The hardware simulations were run on IBM Q 20-qubit Johannesburg on qubit layouts $\left[q_0, q_1\right]=[0,1]$ (points $\beta=0,0.05,0.20,0.30$) and $\left[q_0, q_1\right]=[0,5]$ (points $\beta=0.10,0.15,0.25$). (b) The hardware simulations were run on IBM Q 20-qubit Johannesburg.}
    \label{fig:DeuteronQITEEvsbeta}
\end{figure*}

Using the QITE algorithm we were able to calculate the ground state energy of deuteron for both $N=2$ and $N=3$ cases. Fig.~\ref{fig:DeuteronQITEEvsbeta} depicts the convergence to the ground state energy for $N=2$ and $N=3$ deuteron Hamiltonian.



\begin{figure}[ht!]
	\begin{center}
	\includegraphics[scale=0.33]{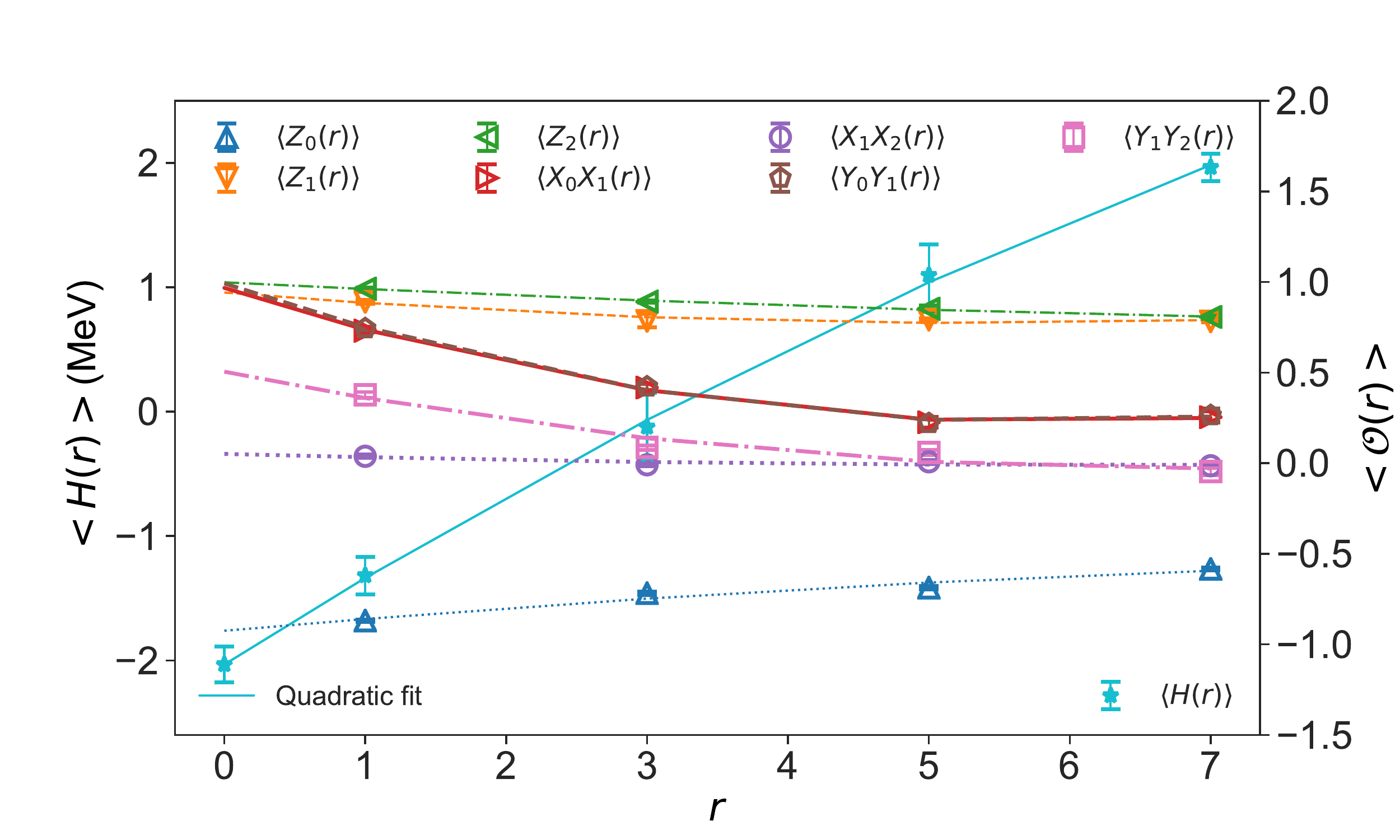}
	\end{center}
\caption{(Color online) Richardson extrapolation of the expectation values of the Pauli operators (on the right axis) and Hamiltonian operator (on the left axis) from their plots as a function of the CNOT gates corresponding to each CNOT gate in quantum circuit in Fig. \ref{fig:DeuteronQCircuit2}(b) for N=3 qubit Hamiltonian at $\beta=0.30$. This simulation was run on IBM Q 20-qubit Johannesburg hardware using the qubit layout $\left[q_0, q_1, q_2\right
]=\left[8, 7, 9 \right]$.
}\label{fig:N3Beta0p03OvsrJohannesburg}
\end{figure}

Data in Fig.s \ref{fig:DeuteronQITEEvsbeta}, \ref{fig:N3Beta0p03OvsrJohannesburg} were obtained after 10 runs each with 8192 shot on IBM Q Johannesburg hardware.
Fig.~\ref{fig:N3Beta0p03OvsrJohannesburg} shows the application of the Richardson extrapolation for $N=3$ case at $\beta=0.30$.
In this figure, the expectation value of the ground state energy and the operators are plotted as a function of the number of CNOT gates corresponding to each CNOT gate in the original quantum circuit.
As a result of our QITE computation the ground state energy for $N=2$ ($N=3$) case is calculated as $E_2=-1.762 \pm 0.2
$ ($E_3=-2.033 \pm 0.1
 $ )
 MeV which is off by $0.76 \%$ ($0.64\%$) from its value obtained from exact diagonalization, i.e. $E_{2, \text{exact}}=-1.749$ MeV ($E_{3, \text{exact}}=-2.046$ MeV). To produce our energy estimates in Fig.~\ref{fig:DeuteronQITEEvsbeta}(a) we sampled several collections of qubits on the chip and used the best results from each set. Readout error mitigation suffice for $\beta=0$ data points for both $N=2$ and $N=3$ case since they don't involve any CNOT gates. To obtain the energy measurements in Fig.~\ref{fig:DeuteronQITEEvsbeta}(a) only readout error mitigation was conducted, except for $\beta=0.30$ where both readout and extrapolation were used. Each experimental point in Fig.~\ref{fig:DeuteronQITEEvsbeta}(b) is the result of post processing with readout error mitigation and Richardson extrapolation.

\textit{Molecular Hydrogen}

\begin{figure*}
    \centering
    \subfloat[]{\includegraphics[width=0.5\linewidth]{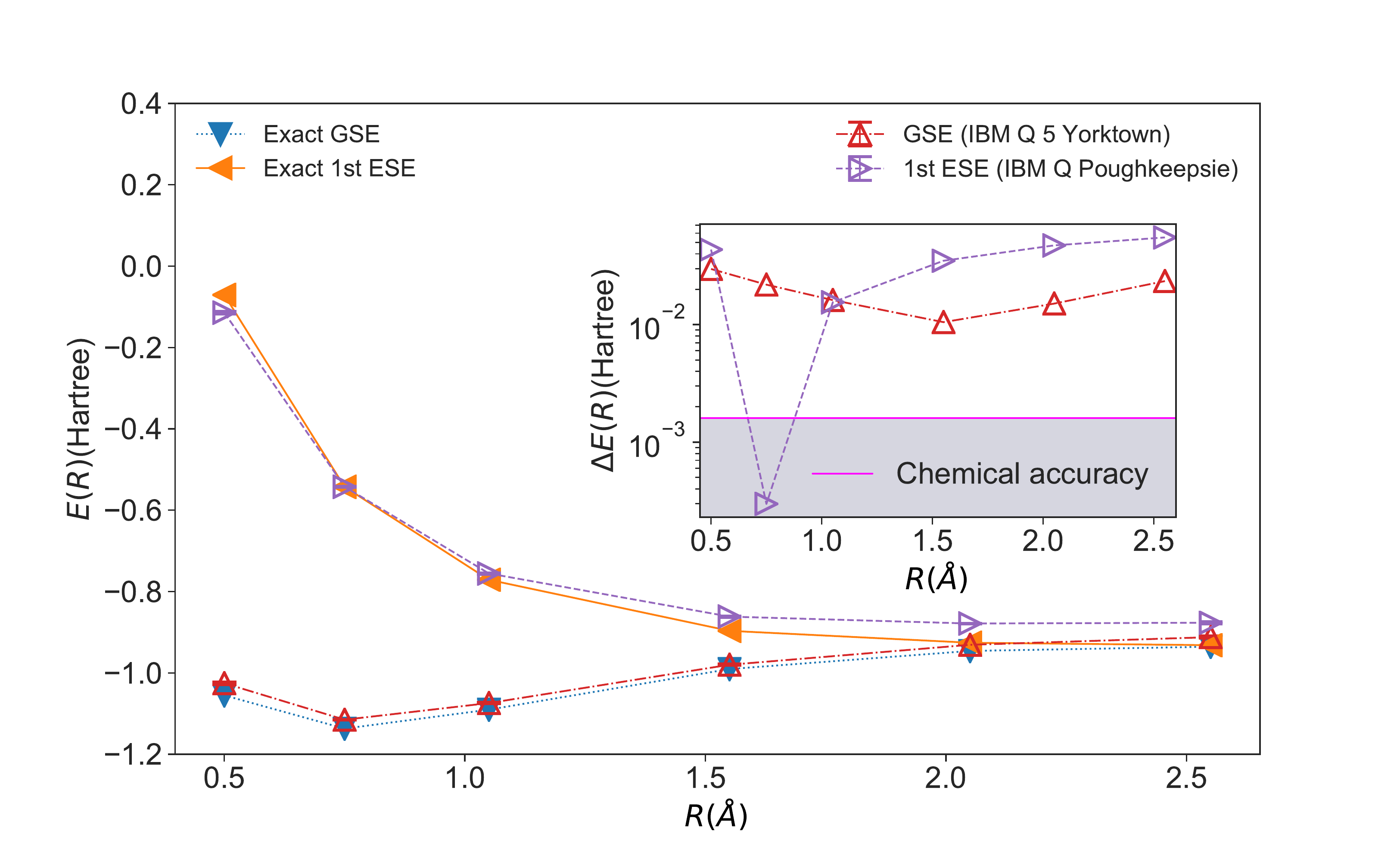}} 
    \subfloat[]{\includegraphics[width=0.5\linewidth]{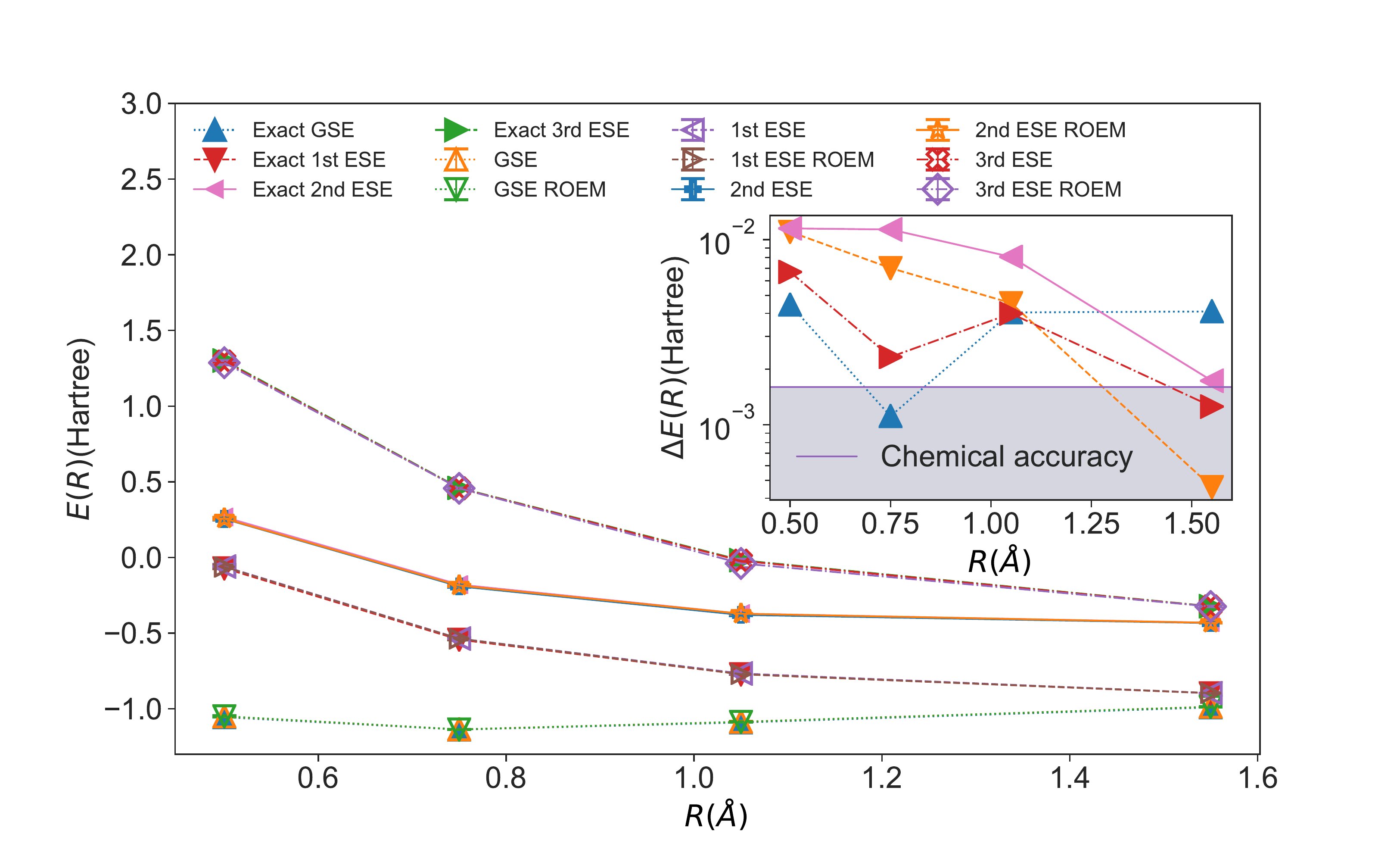}} 
    \caption{(Color online) Energy expectation values of two-qubit molecular Hydrogen as a function of bond length, $R$. We compared the values from exact diagonalization with the values obtained from hardware. The inset shows the relative errors of the quantum computed energy values compared to chemical accuracy. (a) The ground state energy (GSE)  calculations (with $|\Psi_0\rangle=|00\rangle$) were done on IBM Q 5 Yorktown and the first excited state energy (1st ESE) calculations (with $|\Psi_0\rangle=|10\rangle$) were done on IBM Q Poughkeepsie hardware using the QITE algorithm. ROEM and Richardson extrapolation were applied. (b) The GSE and third excited state energy (3rd ESE) calculations (with $|\Psi_0\rangle=|00\rangle$) were done on IBM Q Rochester and the first and second excited state energy (2nd ESE) calculations (with $|\Psi_0\rangle=|01\rangle$) were done on IBM Q Poughkeepsie hardware using the QLanczos algorithm. The values with and without ROEM are presented.}
    \label{fig:2QubitH2QITEandQLanczosEvsR}
\end{figure*}

Although we would expect QITE algorithm to converge to the ground state energy only, we found that depending on the choice of the initial state, $|\Psi_0\rangle$, the excited state energy of the system can also be calculated. In Fig. \ref{fig:2QubitH2QITEandQLanczosEvsR}(a) we plotted the ground and first excited state energies as a function of bond length, $R$, that are obtained using QITE on hardware and compared with the values obtained from exact diagonalization. Because of the availability of devices we used two separate processors for calculation of the ground (on IBM Q 5 Yorktown) and first excited (on IBM Q Poughkeepsie) state energies. The ground (first excited) state energy values are calculated with an initial state of choice $|\Psi_0\rangle=|00\rangle$ ($|\Psi_0\rangle=|10\rangle$). In the case of chemical systems we would like to calculate energy values within chemical accuracy which, is $1.6 \times 10^{-3}$ Hartree. Therefore, in the inset of Fig. \ref{fig:2QubitH2QITEandQLanczosEvsR}(a) we show the relative error in energy ($\Delta E(R)$) as a function of bond length compared with chemical accuracy. QITE was able to obtain chemical accuracy for one or two steps depending on the trial state.

\subsubsection{QLanczos Results}

As mentioned earlier, QLanczos can also be used for quantum computation of both the ground and excited state energies. The choice of the initial state, $|\Psi_0\rangle$, is the one that determines which energies are being calculated. Here, we present our quantum computation of the ground (for deuteron and molecular Hydrogen) and excited state energies (for molecular Hydrogen only - note that the deuteron does not have a bound excited state) using QLanczos.

Quantum computation of the ground and excited state energies using QLanczos might require stabilization of the algorithm as the generalized eigenvalue equation in \eqref{gen_eig} might be numerically ill-conditioned. In our particular deuteron problem, due to the linear dependence of the vectors, $|\Phi_l\rangle$, in Krylov subspace, we had to  perform the stabilization process explained in the supplementary information of \cite{Motta2019}.


We ran QLanczos on two different devices: IBM Q 20-qubit Poughkeepsie (for $N=2$ deuteron and first and second excited state energies of molecular Hydrogen) and IBM Q 53-qubit Rochester (for $N=3$ deuteron and ground and third excited state energies molecular Hydrogen). The statistical error is calculated for $N_{\text{runs}}=5$ for deuteron and $N_{\text{runs}}=3$ for molecular Hydrogen, each run having 8192 shots. Results of our quantum computation of the ground state energies for $N=2$ and $N=3$ deuteron Hamiltonian are summarized in Table \ref{table:QLanczosDeuteron}.
\begin{table}[]
\begin{tabular}{|c|c|c|c|}
\hline
\multicolumn{4}{|c|}{\textit{\textbf{$E$ from exact diagonalization}}}                                       \\ \hline
\multicolumn{2}{|c|}{$N=2$}                            & \multicolumn{2}{c|}{$N=3$}                          \\ \hline
\multicolumn{2}{|c|}{-1.749}                           & \multicolumn{2}{c|}{-2.046}                         \\ \hline
\multicolumn{4}{|c|}{\textit{\textbf{QLanczos $E$ from eigenvalues of \eqref{gen_eig}}}} \\ \hline
\multicolumn{2}{|c|}{$N=2$}                            & \multicolumn{2}{c|}{$N=3$}                          \\ \hline
\textit{Raw}              & \textit{ROEM}             & \textit{Raw}            & \textit{ROEM}            \\ \hline
$-1.024\pm 0.1 $               & $-1.631\pm 0.1 $                & $2.347\pm 0.4 $              & $-1.402\pm 0.5 $               \\ \hline
\multicolumn{4}{|c|}{\textit{\textbf{QLanczos $E$ from \eqref{QLancEn}}}}                 \\ \hline
\multicolumn{2}{|c|}{$N=2$}                            & \multicolumn{2}{c|}{$N=3$}                          \\ \hline
\textit{Raw}              & \textit{ROEM}             & \textit{Raw}            & \textit{ROEM}            \\ \hline
$-1.726 \pm 0.02  $             & $-1.728 \pm 0.02$                & $-2.025 \pm 0.02 $            & $-2.022 \pm 0.02 $              \\ \hline
\end{tabular}
\caption{$N=2$ and $N=3$ ground state energies (in MeV) calculated using the QLanczos algorithm. We ran the simulations on IBM Q 20-qubit Poughkeepsie ($N=2$) and 53-qubit Rochester ($N=3$) hardware. We chose the initial state $|\Psi_0\rangle=|10\rangle$ ($|\Psi_0\rangle=|100\rangle$) for $N=2$ ($N=3$).}
\label{table:QLanczosDeuteron}
\end{table}

In Table \ref{table:QLanczosDeuteron} and Fig.~\ref{fig:2QubitH2QITEandQLanczosEvsR}(b) we present the results for QLanczos with and without readout error mitigation (indicated as ROEM) for the deuteron and molecular Hydrogen, repsectively. The results obtained using \eqref{QLancEn} are in good agreement with the values obtained from exact diagonalization, while energies obtained from the stabilized generalized eigenvalue equation do not agree well with the exact values due to stability issues in the case of molecular Hydrogen. Choosing a smaller regularization parameter would make these values closer to the exact values with a cost of adding more vectors to the Krylov subspace. In our example, a Krylov subspace with two vectors out of $\{|\Phi_0\rangle,|\Phi_2\rangle, |\Phi_4\rangle\}$ subspace were sufficient to obtain the ground and excited state energies for the deuteron and molecular Hydrogen examples. 
For molecular Hydrogen, we used two different initial states ($|\Psi_0\rangle=|00\rangle$ and $|\Psi_0\rangle=|10\rangle$) which helped us to calculate the energy spectrum as a function of the bond length, $R$. 

We found that using \eqref{QLancEn} gives very close values to exact diagonalization with or without readout error mitigation, meaning that QLanczos is potentially noise resilient. Combined with fast convergence the algorithm has a few advantages that make it useful for quantum computation of the ground and excited state energies of many-body systems. Since our QLanczos results are in good agreement with the exact values from diagonalization, we did not perform Richardson extrapolation. This would require 3 more measurements at every QITE step to build the Krylov space.  

Although the computational limits of the quantum computers require us to truncate the harmonic oscillator basis, different schemes were proposed for extrapolating the bound state energies to infinite basis. We will follow the scheme that is based on the 
L\"uscher's formula \cite{Furnstahl2014} that was used in \cite{Dumitrescu2018}.
The extrapolation of the bound state energy values to the infinite basis is listed in Table \ref{table:Luscher}. More information on the extrapolation of the ground state energy to the infinite harmonic oscillator basis can be found in the supplementary material in appendix \ref{sec:supp}.
 
\begin{table}
\centering

\begin{tabular}{c|c|c|c|c|c|l}
\cline{2-6}
&$N$ & $E_N$ & $\mathcal{O}(e^{-2kl})$ & $\mathcal{O}(kLe^{-4kl})$ & $\mathcal{O}(e^{-4kl})$ \\ \cline{1-6}
\multicolumn{1}{ |c  }{\multirow{2}{*}{Exact } } &
\multicolumn{1}{ |c| }{$2$} & -1.749 & -2.394 & -2.194 &  &     \\ \cline{2-6}
\multicolumn{1}{ |c  }{}                        &
\multicolumn{1}{ |c| }{$3$} & -2.046 & -2.336 & -2.199 & -2.209 &     \\ \cline{1-6}
\multicolumn{1}{ |c  }{\multirow{2}{*}{QITE} } &
\multicolumn{1}{ |c| }{$2$} & -1.762 & -2.410 & -2.208 &   \\ \cline{2-6}
\multicolumn{1}{ |c  }{}                        &
\multicolumn{1}{ |c| }{$3$} & -2.033 & -2.334 & -2.198 & -2.174  \\ \cline{1-6}
\multicolumn{1}{ |c  }{\multirow{2}{*}{QLanczos} } &
\multicolumn{1}{ |c| }{$2$} & -1.728 & -2.369 & -2.171 &  \\ \cline{2-6}
\multicolumn{1}{ |c  }{}                        &
\multicolumn{1}{ |c| }{$3$} & -2.022 & -2.311 &  -2.175 & -2.185 \\ \cline{1-6}
\end{tabular}
\caption{L\"uscher's extrapolation of the deuteron bound state energies (in MeV) to the infinite basis.}
\label{table:Luscher}
\end{table}


\section{Conclusion}

In this study, we presented a practical alternative for calculation of the ground and excited state energies of the many-body systems by using \textit{single-step} version of the QITE and QLanczos algorithms presented in \cite{Motta2019} using deuteron and molecular Hydrogen as specific examples. 
This approach may be a good low-depth circuit alternative to other contemporary methods. We also noted that QITE can be used to calculate the excited state energy whose eigenvector is non-orthogonal to the initial state $|\Psi_0\rangle$. We also presented examples of the applications of readout error mitigation and Richardson extrapolation with these algorithms. On the other hand, QLanczos gave results that are good agreement with the exact diagonalization calculations therefore, it did not require additional error mitigation procedures.

We obtained the bound state energy of the deuteron at the next-to-leading order with a 0.5$\%$ (0.9$\%$) error for $N=2$ ($N=3$) using QITE and with a 2.2$\%$ (1.6$\%$) error for $N=2$ ($N=3$) case using QLanczos, compared to its experimental value of $-2.22$ MeV. We also showed the ground and excited state energies of the two-qubit molecular Hydrogen can be calculated within chemical accuracy using the QLanczos algorithm for a few bond lengths.  
\label{sec:con}

\acknowledgments
We acknowledge useful discussions with C. W. Johnson, T. Morris, and E. Dumitrescu. The quantum circuits were drawn using Q-circuit package \cite{QCircuit}.
This work was supported by the Quantum Information Science Enabled Discovery (QuantISED) for High Energy Physics program at ORNL under FWP number ERKAP61 and used resources of Oak Ridge Leadership Computing Facility located at ORNL, which is supported by the Office of Science of the Department of Energy under contract No. DE-AC05-00OR22725.
 The authors acknowledge use of the IBM Q for this work. The views expressed are those of the authors and do not reflect the official policy or position of IBM or the IBM Q team.

\appendix

\section{The model}\label{sec:mod}

We will apply QITE and QLanczos algorithms into two nontrivial systems, i.e. deuteron \ref{sec:Deuteron} and Hydrogen molecule \ref{sec:H2}.

\subsection{Deuteron} \label{sec:Deuteron}
We follow the ref.s \cite{Dumitrescu2018} and \cite{Shehab2019}, in which the pion-less effective field theory (EFT) is implemented through a discrete variable representation in the harmonic oscillator (HO) basis based on ref.s \cite{Binder2016} and \cite{Bansal2017}. Then the pion-less EFT Hamiltonian of the deuteron in the discrete variable representation using the HO basis can be expressed as
\begin{equation}
H_N = \sum_{n,n'=0}^{N-1}\langle n' | (T+V)|n\rangle a_{n'}^\dagger a_n~,
\end{equation}
where $N$ is the maximum number of oscillator quanta included in the HO basis and $a_n$ and $a_n^\dagger$ are, respectively, the annihilation and creation operators for $n=0,1,\dots, N-1$ and they obey fermionic anti-commutation relations
\be
\begin{split}
    &\{a_n,a_{n'}\}=\{a_n^\dagger,a_{n'}^\dagger\}=0
    \\& \{a_n,a_{n'}^\dagger\}=a_n a_{n'}^\dagger+a_{n'}^\dagger a_n=\delta_{n,n'}~.
\end{split}
\ee
The kinetic and potential energy terms in this Hamiltonian can be written as
\begin{equation}
\begin{split}
\langle n'|T|n\rangle&=\frac{\hbar \omega}{2}\Big[(2n+3/2)\delta_n^{n'}-\sqrt{n(n+1/2)}\delta_{n}^{n'+1}\\&\, \, \, \ \ \ \ \ \ \ \ -\sqrt{(n+1)(n+3/2)}\delta_n^{n'-1}\Big]~,\\
\langle n'|V|n\rangle&=V_0\delta_n^0 \delta_n^{n'}~.
\end{split}
\end{equation}
We choose the HO energy spacing as $\hbar \omega \approx 7$ MeV, the potential coefficient as $V_0\approx -5.686$ MeV and the ultraviolet cutoff for the potential as $\Lambda\approx 152$ MeV .

The simulation of the physical systems on quantum computers is made possible by mapping the creation and annihilation operators onto Pauli matrices. This process is done using the Jordan-Wigner transformation \cite{Jordan1993} and for $N=2$ and 3 we obtain 
\begin{equation}
\begin{split}
    H_2 &= 5.907 I + 0.2183 Z_0 \\ 
    &\ \ \ - 6.125  Z_1-2.143(X_0X_1+Y_0Y_1)\\
    H_3 &= H_2+9.625(I-Z_2)-3.913(X_1X_2+Y_1Y_2)~,
\end{split} \label{DeutHam}
\end{equation} 
with the Pauli matrices defined as
\be
\sigma_j=I \otimes \dots \otimes \sigma \otimes \dots \otimes I \label{eq:Pauli}
\ee
where $\sigma\in \{X, Y, Z\}$ is in the $j$-th position with $j=0,\dots, N-1$, $\otimes$ indicates tensor product and $I$ is the identity matrix.

\subsection{Hydrogen Molecule} \label{sec:H2}

We will use the two-qubit molecular Hydrogen Hamiltonian \cite{OMalley2016}
\begin{equation}
\begin{split}
    H(R) &= h_0(R) I+h_1(R) Z_0+h_2(R) Z_1 + h_3(R) Z_0 Z_1 \\& + h_4(R) X_0 X_1 +h_5(R) Y_0 Y_1~,
\end{split}
\end{equation} 
where coefficients $h_i(R)$ for $i\in \{0,1, \dots, 5\}$ are real-valued and functions of bond length, $R$, of the molecule. For calculation of the binding and excited state energies of the Hydrogen molecule we will use the coefficients calculated in STO-3G basis given in Table I of supplementary information of \cite{Colless2017}.

\section{Error Mitigation}
\label{sec:err}

The noise due to the nature of the quantum simulators requires the application of the error mitigation strategies. Although there are various error mitigation strategies
proposed in the literature, for our purposes, we used readout error mitigation and Richardson extrapolation techniques to reduce the noise involved in our calculations.    

Out of the different sources of errors in a quantum circuit the readout errors are the errors associated with the final measurements in the quantum circuit. Therefore, we start by mitigating these errors in our quantum computation. To this end, we use the readout error mitigation scheme proposed in \cite{Yeter2019}. In that scheme, the expectation values of the operators in the Hamiltonian are calculated using the following formula.

\be
\begin{split}
\label{eq:roem}
\expect{\sigma_i \dots \sigma_j}=&\sum_{ x \in {\text{possible outcomes}}}p(x)\\& \ \ \ \ \ \times\frac{(-1)^{x_{i}}-p_i^-}{1-p_i^+}\times\dots \times\frac{(-1)^{x_{j}}-p_j^-}{1-p_j^+}~,
\end{split}
\ee
where $p(x)$ is the probability of each qubit outcome and it takes $2^{N}$ values. For example, for $N=2$, $x \in \{00,01,10,11\}$. The symmetric and anti-symmetric combinations of the probability of $i$-th qubit flipping from 0 to 1 ($p_i(0|1)$) or from 1 to 0 ($p_i(1|0)$) is defined as
\be
p_i^\pm=p_i(0|1)\pm p_i(1|0)~.
\ee
Although $p_i(0|1)$ and $p_i(1|0)$ values are provided by IBM's Qiskit library, to get the most up-to-date values we obtained the readout error probabilities by preparing each qubit in computational basis 0 and 1 and then performing a measurement on each qubit in each case which gives us
\be
p(1|0)= \frac{\# \ \text{of states prepared in $|1\rangle$ measured in $|0\rangle$}}{\# \ \text{of shots}}
\ee
or vice versa for $p(0|1)$. We ran the simulations using 8192 number of shots. To propagate the error due to the statistical error in the readout errors for $N=2$ deuteron case we did the readout error measurements 10 times and propagated the statistical error in measurements and statistical error in readout measurements in our results. As a result of our experimental measurements the statistical error in measurements is not different than the statistical error in readout error measurements therefore, we calculated the statistical error only for our $N=3$ deuteron and molecular Hydrogen calculations.

Although we were able to reduce the depth of the quantum circuit using the \textit{single-step} method, the decoherence effects became apparent in the expectation value measurements. Therefore, in addition to the readout error mitigation we also used the Richardson extrapolation (\cite{Li2017}, \cite{Kandala2018}) technique for the short-depth quantum circuits \cite{Temme2017}
to mitigate the errors associated with the noise produced by the gates used in the quantum circuit. The basic idea in this technique is to increase the error rate deliberately by a constant factor of $r$ which is followed by an extrapolation to obtain the noise free expectation value. In this particular study, we increase the error rate by adding pairs of CNOT gates. The process of adding CNOT pairs is not expected to change the result of measurements since it corresponds to an identity matrix but it will contribute to the noise produced by CNOT gates. Our results showed that for two-qubit systems the expectation values of the observables scale linearly as
\be
\expect{\mathcal{O}(r)}=A r+ \expect{\mathcal{O}(0)}
\ee
and for $N=3$ deuteron system they scale quadratically as
\be
\expect{\mathcal{O}(r)}=A r^2+B r+ \expect{\mathcal{O}(0)}
\ee
where the coefficients $A$, $B$, and the extrapolated noiseless expectation value $\expect{\mathcal{O}(0)}$ are found from the linear and quadratic fit to the data points of the expectation values of the operators for each case. We did not apply Richardson extrapolation technique to the QLanczos measurements since the results obtained using the QLanczos algorithm were in good agreement with the exact diagonalization results.

\section{Extrapolation to the infinite harmonic oscillator basis}
\label{sec:supp}
\begin{table*}[ht!]
\begin{tabular}{|c|c|c|}
\hline
\textbf{Variable}                           & \textbf{Symbol, Equation}                  & \textbf{Value}       \\ \hline
finite-basis energy                         & $E_N$                                      &                      \\ \hline
infinite-basis energy                       & $E_\infty=-\frac{\hbar^2k_\infty^2}{2\mu}$ &                      \\ \hline
binding momentum                            & $k_\infty$                                 &                      \\ \hline
reduced mass                                & $\mu=\frac{m_p+m_n}{4}$                    & 469.45925 MeV/$c^2$  \\ \hline
proton mass                                 & $m_p$                                      & 938.272 MeV/$c^2$    \\ \hline
neutron mass                                & $m_n$                                      & 939.565 MeV/$c^2$    \\ \hline
\multirow{3}{*}{effective hard-wall radius} & \multirow{3}{*}{$L(N)$}                    & $L(1)=9.14$ fm       \\ \cline{3-3} 
                                            &                                            & $L(2)=11.45$ fm      \\ \cline{3-3} 
                                            &                                            & $L(3)=13.38$ fm      \\ \hline
conversion constant                         & $\hbar c$                                  & 197.326 MeV$\cdot$fm \\ \hline
energy spacing                              & $\hbar\omega$                              & 7 MeV                \\ \hline
\end{tabular}             
\caption{The values and definitions of the variables in \eqref{eq:Luscher}. }
\label{table:LuscherSupp}
\end{table*}
The finite-size corrections to the infinite size harmonic oscillator basis based on the L\"uscher's method can be stated as

\be
E_N-E_\infty = \mathcal{A} e^{-2k_\infty L}+\mathcal{B} k_\infty L e^{-4k_\infty L}+\mathcal{C} e^{-4k_\infty L}~,\label{eq:Luscher}
\ee
where
\be
\begin{split}
    &\mathcal{A}=\frac{\hbar^2 k_\infty \gamma^2}{m}~, \ \ \ \mathcal{B}=\frac{2 \hbar^2\gamma^4}{m}~, \\&
    \mathcal{C}=\frac{\hbar^2 k_\infty \gamma^2}{\mu}\left(1-\frac{\gamma^2}{k_\infty}-\frac{\gamma^4}{4k_\infty^2}+2w_2k_\infty\gamma^4\right)~.
\end{split}
\ee

The values and definitions of the variables in \eqref{eq:Luscher} are given in Table \ref{table:LuscherSupp}.
The terms in right-hand side of \eqref{eq:Luscher} refer to leading order (LO), next-to-leading order (NLO) and N2LO, respectively. Curve fitting the LO and NLO terms gives the binding momentum, $k_\infty$ and the asymptotic normalization coefficient, $\gamma$, for each order by using $E_1$ and $E_2$. Fitting to N2LO term adding $E_3$ data helps calculating an effective range parameter, $w_2$.



\section{Information on Experimental Runs on IBM Q Hardware}
\begin{table}[h!]
\begin{tabular}{|c|c|c|c|}
\hline
\textit{\textbf{\begin{tabular}[c]{@{}c@{}}Figure\\ / Table\end{tabular}}} & \textit{\textbf{$\#$ of shots}} & \textit{\textbf{$\#$ of runs}} & \textit{\textbf{\begin{tabular}[c]{@{}c@{}}IBM Q \\ hardware\end{tabular}}}              \\ \hline
Fig. \ref{fig:DeuteronQITEEvsbeta}(a)                                                                     & 8192                            & 10                             & Johannesburg (v1.1.5)                                                                   \\ \hline
Fig. \ref{fig:DeuteronQITEEvsbeta}(b)                                                                     & 8192                            & 10                             & Johannesburg (v1.1.5)                                                                   \\ \hline
Fig. \ref{fig:N3Beta0p03OvsrJohannesburg}                                                                     & 8192                            & 10                             & Johannesburg  (v1.1.5)                                                                  \\ \hline
Table \ref{table:QLanczosDeuteron} (N=2)                                                              & 8192                            & 5                              & Poughkeepsie (v1.2.6)                                                                   \\ \hline
Table \ref{table:QLanczosDeuteron} (N=3)                                                              & 8912                            & 5                              & Rochester (v1.1.1)                                                                      \\ \hline
Fig. \ref{fig:2QubitH2QITEandQLanczosEvsR} (a)                                                                 & 8192                            & 3                              & 5 Yorktown (v2.0.1)                                                                        \\ \hline
Fig. \ref{fig:2QubitH2QITEandQLanczosEvsR} (b)                                                                 & 8192                            & 3                              & \begin{tabular}[c]{@{}c@{}}Poughkeepsie \\ and Rochester\end{tabular} \\ \hline
\end{tabular}
\caption{Information about the experimental runs on hardware.}
\label{table:Hardware}
\end{table}

%

\end{document}